\documentclass[fleqn,12pt,twoside]{article}
\usepackage{espcrc1}

\usepackage{graphicx}
\usepackage[figuresright]{rotating}

\newcommand{\AmS}{{\protect\the\textfont2
  A\kern-.1667em\lower.5ex\hbox{M}\kern-.125emS}}

\title{Branching ratio measurements  of the 7.12-MeV state in 
$^{16}$O}

\author{C. Matei\address{Edwards Accelerator Laboratory,
        Department of Physics and Astronomy, \\
	Ohio University, Athens OH 45701, USA}
	and 
        C. R. Brune\addressmark}

\begin{document}

\maketitle

\begin{abstract}
Knowledge of the $\gamma$-ray branching ratios of the 7.12-MeV state 
of $^{16}$O is important for the extrapolation of the 
$^{12}$C($\alpha$,$\gamma$)$^{16}$O cross section to astrophysical energies.
Ground state transitions provide most of the $^{12}$C($\alpha$,$\gamma$)$^{16}$O
total cross section while cascade transitions have contributions of the 
order of 10-20$\%$. Determining  the $7.12\rightarrow 6.92$-MeV branching ratio
will also result in a better extrapolation of the cascade and $E2$ 
ground state cross section to low energies. We report here on measurements on 
the branching ratio of the $7.12\rightarrow 6.92$-MeV transition in $^{16}$O. 
\end{abstract}

\section{INTRODUCTION}

A typical star spends 90$\%$ of its lifetime burning hydrogen on the Main 
Sequence. When hydrogen is exhausted in the center, the remaining helium 
core contracts transforming gravitational potential energy into thermal 
energy. As the temperature increases helium burning starts. The first 
step of this process is the triple-$\alpha$ capture to form $^{12}$C,
followed by the $^{12}$C radiative capture of $\alpha$ particles to 
form $^{16}$O \cite{Wa}. In the helium-burning phase of stellar evolution the 
triple-$\alpha$ process and $^{12}$C($\alpha$,$\gamma$)$^{16}$O are 
the most important reactions. Their relative reaction rates determine the 
$^{12}$C$\slash$$^{16}$O ratio at the end of helium burning and beyond.
The triple-$\alpha$ reaction rate is well known for stellar evolution 
calculations. However the rate of $^{12}$C($\alpha$,$\gamma$)$^{16}$O 
is poorly determined as the cross section was measured experimentally
down to energies around 1 MeV and needs to be extrapolated to 
helium-burning energies around 300 keV.
The $^{12}$C($\alpha$,$\gamma$)$^{16}$O cross section
is composed of 3 major parts: E1, E2 ground state transitions and
cascade transitions. Ground state transitions provide most 
of the total cross section while cascade transitions have contributions 
of the order of 10-20$\%$.  Determining the $7.12\rightarrow6.92$-MeV
branching ratio  will result in a better extrapolation of the cascade 
and $E2$ ground state cross sections to lower energies \cite{Bu}.

\section{EXPERIMENTAL DETAIL}

\begin{figure}[t]
\begin{minipage}[t]{77mm}
\includegraphics[scale=0.95]{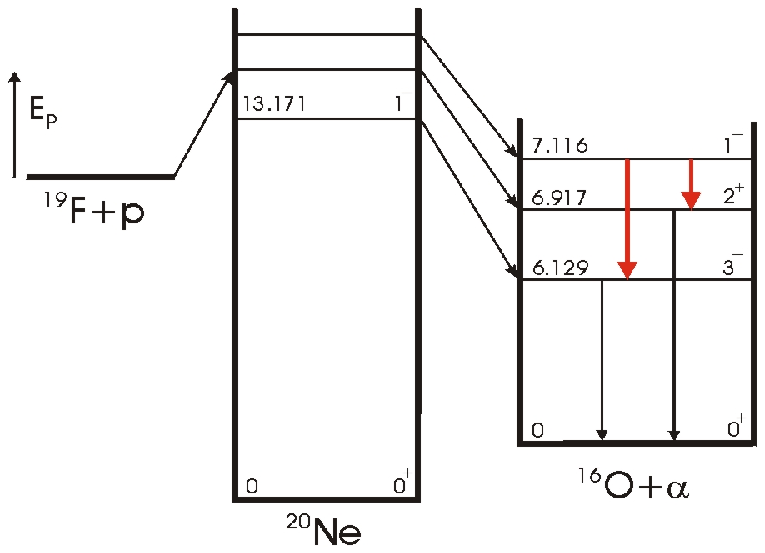}
\caption{Level schemes of $^{16}$O and $^{20}$Ne to illustrate how 
the $^{19}$F(p,$\alpha$$\gamma$)$^{16}$O reaction proceeds.}
\label{fig:levels}
\end{minipage}
\hspace{\fill}
\begin{minipage}[t]{75mm}
\includegraphics[scale=0.42]{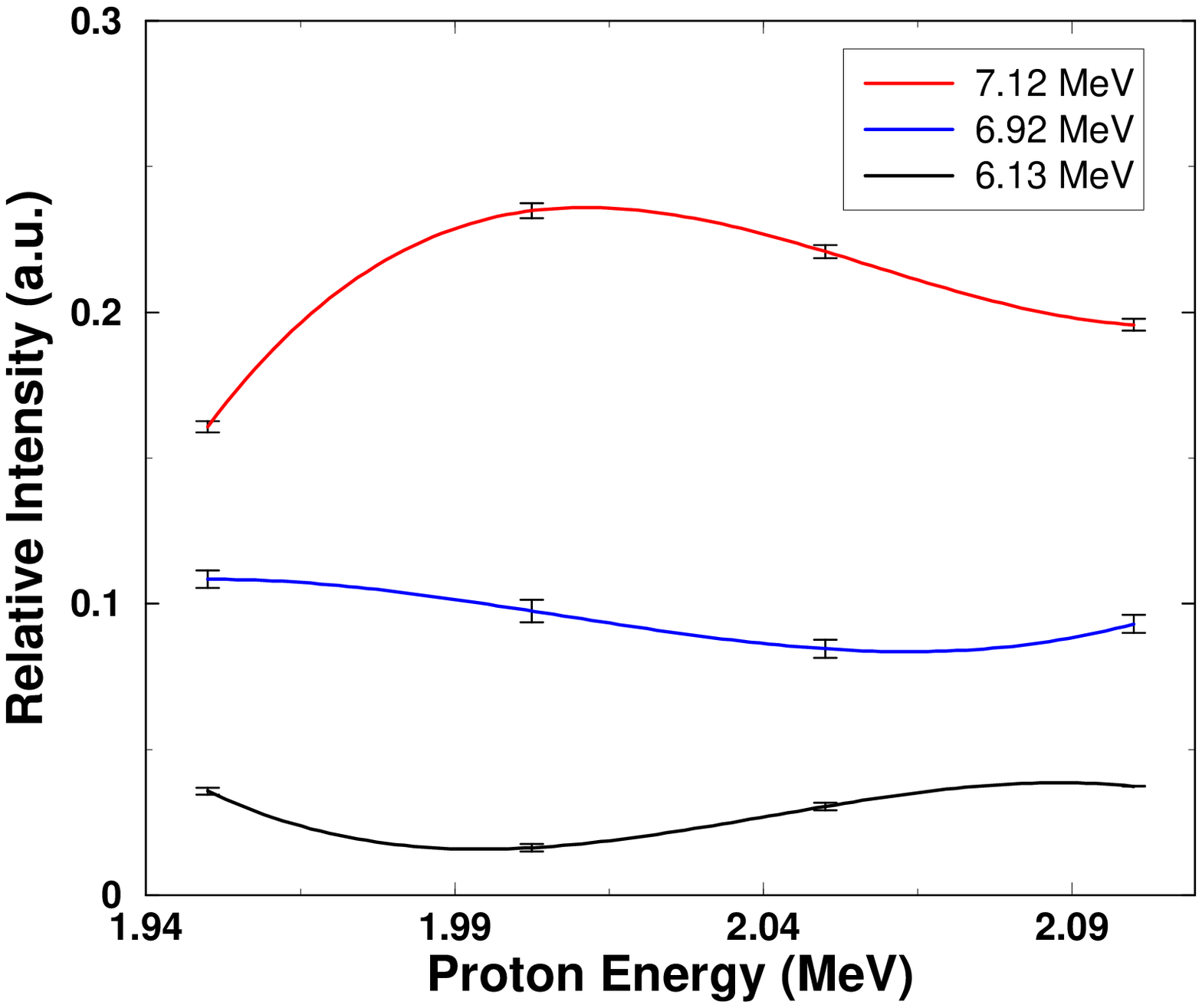}
\caption{Relative intensities of the three transitions at 
proton energies between 1.95 MeV and 2.1 MeV}
\label{fig:peakratio}
\end{minipage}
\end{figure}

The 7.12-MeV excited state in $^{16}$O is formed via the 
$^{19}$F(p,$\alpha$)$^{16}$O$^*$ reaction; a level diagram
is shown in Fig.~\ref{fig:levels}.
A 100-$\mu$g/cm$^{2}$-thick CaF$_{2}$ layer evaporated onto
a 1-mm-thick carbon backing was bombarded by
protons produced at the 4.5-MV Tandem Van de Graaf
Accelerator at Ohio University.

An angular distribution study for the 7.12-MeV $\gamma$-ray at 4 
different proton energies was performed. These measurements offered 
a better understanding of the $\gamma$-ray angular distribution for
the strong transitions and information on the dependence of the 
relative intensity of the three $\gamma$-ray components with 
proton energy. In Fig.~\ref{fig:peakratio} the relative intensities 
of the strong transitions in the direction of the beam are shown.
For the branching-ratio measurements the proton beam energy
was chosen to be E$_{p}$=2.0025 MeV in order
to maximize the relative population of the 7.12-MeV state \cite{Ask}. 

\begin{figure}[h]
\center
\includegraphics[width=10.8cm,height=5.7cm]{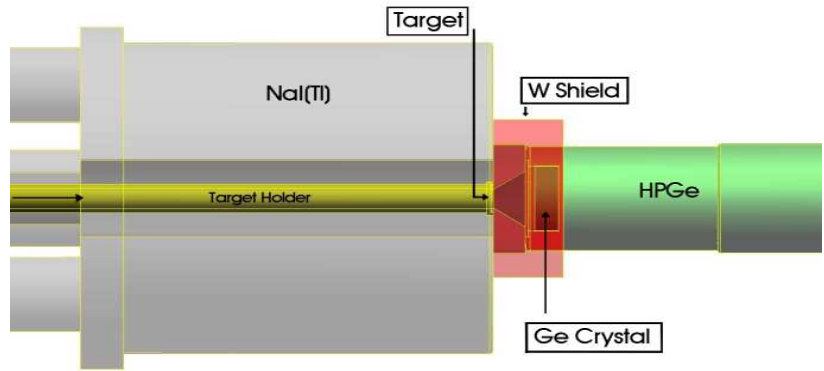}
\caption{Schematic view of the experimental setup. The beam enters
from left.}
\label{fig:upview} 
\end{figure}

The coincidence setup uses a two-detector configuration. For the 
detection of the $6.92\rightarrow 0$-MeV transition,
a 22.9-cm~$\times$~22.9-cm NaI(Tl) split annulus 
consisting of two optically-isolated sections is used. The large 
volume of the crystal offers a 30$\%$ efficiency in detecting 
the 6-7 MeV $\gamma$-rays. The inside cylindrical hole in the 
crystal is able to accommodate an extended target holder as 
sketched in Fig.~\ref{fig:upview}.

For the $7.12\rightarrow 6.92$-MeV transition a HPGe detector is used.
This detector has a planar geometry to lower the noise and achieve better
time resolution for low-energy gamma rays.
The efficiency of different detector geometries and thicknesses was 
simulated using the GEANT3 simulation package. The simulation looked 
for the optimum Ge thickness to maximize the photopeak efficiency 
at 200 keV while minimizing the sensitivity to 7-MeV $\gamma$-rays. 
A planar crystal of 1.5-cm thickness and 2000-mm$^{2}$ area was chosen.  

To further lower the background in the 0.2-MeV region of interest
a tungsten shield was utilized to minimize the scattering of $\gamma$-rays
between detectors. The design of the shield was also guided by
the GEANT3 simulation package.

\section{EXPERIMENTAL RESULTS}

Data on $\gamma$-$\gamma$ coincidences was taken for 4 days, in 24 hour 
shifts, with beam currents ranging from 10 to 20 nA. A standard 
coincidence setup fed the signal from the detectors and the 
Time-to-Amplitude Converter (TAC) into a 4096 channels data acquisition system.
The time resolution of the system (for energy signals of interest)
was 5~ns.

\begin{figure}[h]
\begin{minipage}[h]{160mm}
\center
\includegraphics[scale=0.84]{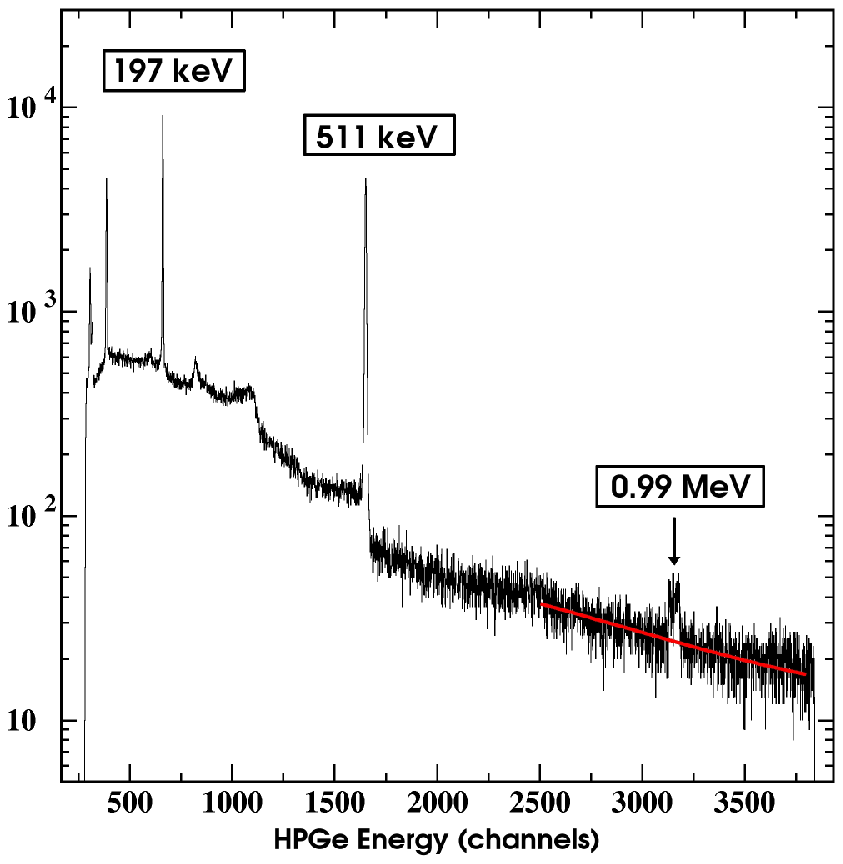}
\hspace{\fill}
\includegraphics[scale=0.84]{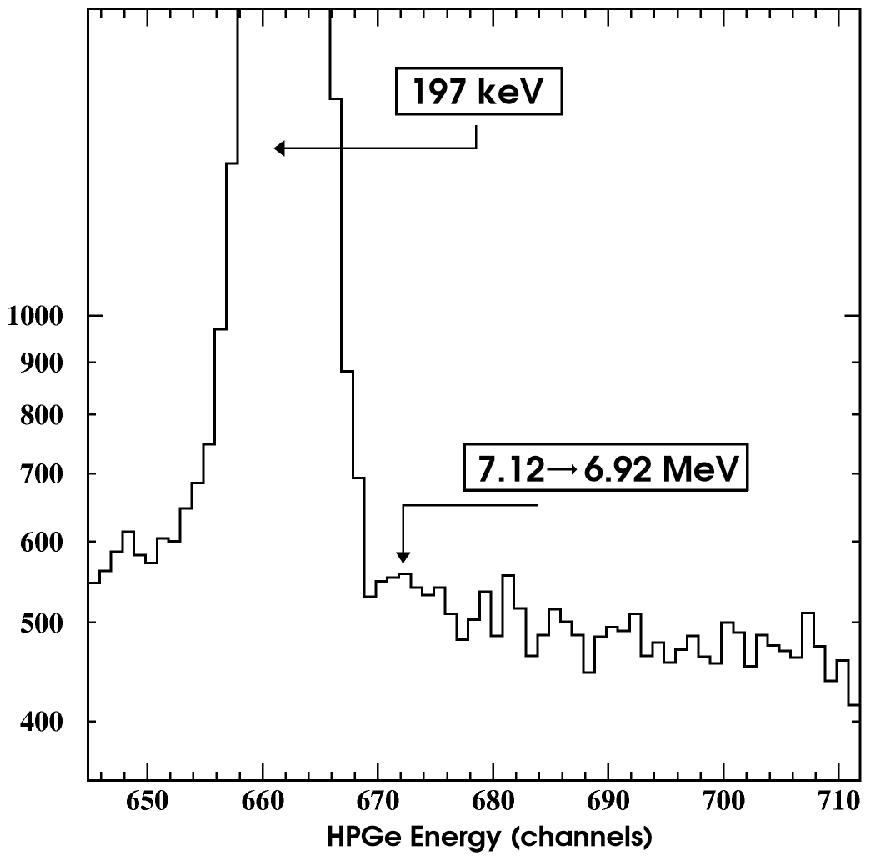}
\caption{HPGe spectrum in coincidence with the 6-7 MeV region in NaI(Tl).
Left figure: The 0.99 MeV peak from the $7.12\rightarrow 6.13$-MeV 
transition is indicated. Right figure: Expanded 200-keV region with 
the 197-keV peak from the $^{19}$F(p,p$^{\prime}$) reaction 
and the position of a possible $7.12\rightarrow 6.92$-MeV transition.}
\label{fig:spectrum}
\end{minipage}
\end{figure}

The spectrum shown in Fig.~\ref{fig:spectrum} was obtained 
by the HPGe detector using gate settings on the TAC coincidence peak 
and the 6-7 MeV region in the NaI(Tl) detector.

For the $7.12\rightarrow 6.13$-MeV transition the peak is clearly
visible between channels 3000 and 3250 in Fig.~\ref{fig:spectrum}. 
The region surrounding the peak was fitted to subtract the 
background and obtain the number of counts in the $7.12\rightarrow 6.13$-MeV 
transition peak. Calibrated sources and GEANT3 simulations were used
to find the efficiencies of the detectors. The result obtained for
this branching ratio is $(8.4\pm 0.4)\times 10 ^{-4}$, in good
agreement with a previous measurement by Wilkinson et al.~\cite{Wi}.

The coincident HPGe spectrum near 200~keV is dominated by the
197.1-keV line which results from random-coincidence
$^{19}$F(p,p$^{\prime}$) events detected in the HPGe.
No peak is evident at 199.8~keV corresponding to the
$7.12\rightarrow 6.92$-MeV transition.
It should be noted that the HPGe detector resolution is 1.3~keV at 200~keV,
so energy range of interest is well resolved from the 197.1-keV line.
From the spectrum shown in Fig.~\ref{fig:spectrum}
we have obtained the following upper limit with a 2-$\sigma$ confidence level:
\begin{equation}
\frac{\Gamma({7.12 \rightarrow 6.92})}{\Gamma({7.12 \rightarrow 0})}
  \leq 1.2\times10^{-5}.
\end{equation}
Note that if the $7.12\rightarrow 6.92$-MeV transition were to have the
same Weisskopf strength as the $7.12\rightarrow 0$-MeV transition
this branching ratio would be $(0.2/7.12)^3=2\times 10^{-5}$.

The results are preliminary and further analysis of the data may reveal 
more information. Histogram manipulation and improved GEANT3 simulations will 
be performed to further understand the limitations of the present setup
and perhaps discover improvements to the experimental approach.

\section{ACKNOWLEDGMENTS}

It is a pleasure to thank J.~Bevington, D.~Carter, C.~Dodson, D.~Jacobs,
T.~Massey, J.~O'Donnell, Y.~Parpottas, A.~Salas, and R.~Wheeler
for assistance with the experiments. This work was supported in part
by the U.S. Department of Energy, under Grant No. DE-FG02-88ER40387.

\end{document}